# CHANNEL ESTIMATION AND MULTIUSER DETECTION IN ASYNCHRONOUS SATELLITE COMMUNICATIONS

Helmi Chaouech[1] and Ridha Bouallegue[2]

[1,2] 6'Tel Research Unit, Higher School of Communications of Tunis (Sup'Com),
University 7th November at Carthage, Tunis, Tunisia
[1,2] National Engineering School of Tunis, Tunisia
[1]helmi.chaouech@planet.tn
[2]ridha.bouallegue@gnet.tn

## ABSTRACT

*In this paper, we propose a new method of channel estimation for asynchronous additive white Gaussian noise channels in satellite communications. This method is based on signals correlation and multiuser interference cancellation which adopts a successive structure. Propagation delays and signals amplitudes are jointly estimated in order to be used for data detection at the receiver. As, a multiuser detector, a single stage successive interference cancellation (SIC) architecture is analyzed and integrated to the channel estimation technique and the whole system is evaluated. The satellite access method adopted is the direct sequence code division multiple access (DS CDMA) one. To evaluate the channel estimation and the detection technique, we have simulated a satellite uplink with an asynchronous multiuser access.*

## 1. INTRODUCTION

In satellite communications, users transmit their data over a wireless support. With use of different spreading sequences such as DS CDMA ones, several users can access the same frequency band simultaneously. In the uplink of DS CDMA wireless systems, the access of users are asynchronous since each one begins to transmit its data independently of others. Thus, each user signal has a specific propagation delay. So, it is not possible to have orthogonal codes during the transmission, which leads to multiple access interference (MAI) occurring. Moreover, transmitted signals are affected by thermal noise and fading produced by the wireless channel. At the receiver, the composite signal, which is formed by active users' data, must be cleaned of channel impairments. Thus, channel estimation is needed to perform equalization task. Detection techniques recover original data from the received signal. Their performances are mainly influenced by the channel estimation. So, in order to obtain high performance of the detectors, we must provide to them good estimated channel parameters. In GEO satellite applications, especially for fixed communications services, ground stations are installed in high places and directed towards the satellite (see figure 1). Thus, additive white Gaussian noise (AWGN) channel is an appropriate model that can describes the wireless channel for these applications. In this paper, we have developed an estimation technique for AWGN channels in asynchronous DS CDMA system. And, we have integrated this channel estimation algorithm in a single stage successive interference cancellation multiuser detector (SIC) in order to evaluate the performance of detection. The channel estimation technique developed is based on signals correlation and MAI suppression. This method allows the estimation of the signal amplitude and the propagation delay of each user. Then, these parameters are fed to multiuser detector to recover the original data transmitted by each user.

The remainder of this paper is organized as follows. Section 2 presents the related work. Section 3 details the DS CDMA access and transmission in asynchronous satellite systems, and presents the conventional detector which is based on a bank of matched filters. Section 4 describes the channel estimation technique and its implementation. Section 5 presents the single stage SIC detector and its channel estimation method integration. Section 6 shows the simulations results and gives some analysis and interpretation of the obtained curves. Finally, in section 7 we have drawn some conclusions from the work.

## 2. RELATED WORK

In the literature, several works deal with channel modelling and estimation problems, and detection techniques. In [2], the authors are focused on channel estimation of a new active user on the system. In [4], channel estimation methods for multipath environment are studied. In [7], other channel and propagation delay estimation algorithms for wireless systems are proposed and their performances are evaluated. In [8], authors have discussed and compared some wireless channel models. In [9] flat Rayleigh fading channel is considered and analysed for image transmission. In [10], time-varying wireless channel is analyzed and evaluated for video transmission. After channel estimation, detections techniques construct the data of each user from the received signal. The conventional detector, which is based on a bank of matched filters, suffers from multiple access interference and is very sensitive to near-far problem [1], [12]. To combat optimally these problems, the maximum likelihood detector is introduced in [3]. This detector has optimal performance under MAI, and near-far resistance at the cost of a computational complexity making its practical implementation not effective. Linear multiuser detectors such as the minimum mean-squared error (MMSE) detector and the decorrelator, deal seriously with the near-far and MAI problems [1], [13], [14]. But, their computation leads to noise enhancement, and needs the inversion of big dimension matrices in the case of asynchronous communications. The subtractive interference cancellation multiuser detectors are good solutions to combat the MAI and resist to near-far effect. In such systems, which operate in some stages, data detection performance is improved from a stage to an other [5], [12], [16], [17], [18]. These detectors are divided into two main architectures; the successive interference cancellation (SIC) detector and the parallel interference cancellation one. In the first architecture, the users' data are dealt serially in each stage, while, in the PIC structure, the detection of each users bit is done simultaneously in each stage. Some iterative detection techniques which don't adopt a CDMA access are proposed in [6], [15]. These methods use an FDMA access and some of them are based on multibeam technology. In [11], [19], channel estimation methods and detection techniques for satellite communications are proposed. These techniques consider a multibeam communication and are based on interference cancellation.

## 3. SATELLITE UPLINK MODEL AND MATCHED FILTER DETECTION

We consider an asynchronous DS CDMA model with K users. The signal of each user corresponds to the transmission of N bits across an AWGN channel.

At the satellite, the received baseband signal is given by:

$$r(t) = \sum_{k=1}^{K} r_k(t) + n(t) \qquad (1)$$

Where $n(t)$ is a zero average additive white Gaussian noise with power $\sigma^2$ and $r_k(t)$ is the k$^{th}$ user signal (ground station signal). It can be written as follows:

$$r_k(t) = \sum_{i=0}^{N-1} A_k b_k^{(i)} s_k(t - iT_b - \tau_k) \qquad (2)$$

Where $A_k$ and $\tau_k$ are the user k signal amplitude and propagation delay respectively. $b_k^{(i)} \in \{+1, -1\}$ is a sequence of antipodal modulated bits which are defined by their period $T_b$.

$s_k(t)$ is the chip sequence which corresponds to the spreading code of user k. It is given by:

$$s_k(t) = \sum_{j=0}^{N_c - 1} c_k^{(j)} \psi(t - jT_c) \qquad (3)$$

Where $\psi(t)$ is a waveform of time during $T_c$ and $c_k^{(i)} \in \{\pm 1/\sqrt{N_c}\}$ is a normalized chip sequence of time duration $N_c T_c$, with $T_c$ is the period of a chip.

For simplicity and without loss of generality, we assume an ordering on the time delays $\tau_k$ such that $0 = \tau_1 \leq \tau_2 \leq \ldots \leq \tau_K < T_b$.

The detections techniques operate at the outputs of the K users matched filters. For a user k, the output of his matched filter which corresponds to the i$^{th}$ bit can be written as:

$$y_k^{(i)} = \int_{-\infty}^{+\infty} r(t) s_k(t - iT_b - \tau_k) dt \qquad (6)$$

The cross-correlation between signatures of users $k$ and $l$ for the i$^{th}$ bit can be defined as:

$$R_{k,l}^{(i)} = \int_{-\infty}^{+\infty} s_l(t - \tau_l) s_k(t - iT_b - \tau_k) dt \qquad (7)$$

From (7), we can draw the following cross-correlation properties [6]:

i. $R_{k,l}^{(i)} = 0 \ \forall i / |i| > 1$ because $s_k(t) = 0 \quad for \ t \notin [0, T_b]$.
ii. $R_{k,l}^{(i)} = R_{l,k}^{(-i)}$ hence $R(-1) = R^T(1)$; where $R(i) = \left(R_{k,l}^{(i)}\right)_{k,l}$, and $(.)^T$ denotes the transpose operator.
iii. $R_{k,k}(0) = 1 \ \forall k$.
iv. For $k>l$, we have $\tau_k > \tau_l$; so, $R_{k,l}^{(1)} = 0$, from where $R(1)$ is an upper triangular matrix with zero diagonal.
v. $\forall k, l \ R_{k,l}^{(0)} = R_{l,k}^{(0)}$, hence, $R(0)$ is symmetric.

Taken into account (1), (2) and (7), the expression (6) can be written as follows:

$$y_k^{(i)} = \sum_{l=1}^{K} \sum_{j=0}^{N-1} A_l R_{k,l}^{(i-j)} b_l^{(j)} + n_k^{(i)} \qquad (8)$$

With use of the cross correlation proprieties, we can express the $k^{th}$ matched filter output for the bit i by:

$$y_k^{(i)} = \sum_{l=1}^{K} A_l \left( R_{k,l}^{(-1)} b_l^{(i+1)} + R_{k,l}^{(0)} b_l^{(i)} + R_{k,l}^{(1)} b_l^{(i-1)} \right) + n_k^{(i)} \tag{9}$$

As we can see from expression (9), it is clear that the decision for the $i^{th}$ bit of user k depends on bits i, i+1 and i-1 of other users.

Generalization of expression (9) for a K users system becomes:

$$Y(i) = R^T(1)AB(i+1) + R(0)AB(i) + R(1)AB(i-1) + \aleph(i) \tag{10}$$

Where: $Y(i) = \left[y_1^{(i)}, y_2^{(i)}, \ldots, y_K^{(i)}\right]^T$, $A = diag[A_1, A_2, \ldots, A_K]$, $B(i) = \left[b_1^{(i)}, b_2^{(i)}, \ldots, b_K^{(i)}\right]^T$ and $\aleph(i) = \left[n_1^{(i)}, n_2^{(i)}, \ldots, n_K^{(i)}\right]^T$.

For the transmission of a finite sequence of N bits, we obtain the following matrix presentation:

$$Y = ZWB + \aleph \tag{12}$$

Where: Z and W are two $NK \times NK$ matrix defined as follows:

$$Z = \begin{bmatrix} R(0) & R(-1) & 0 & \cdots & 0 \\ R(1) & R(0) & R(-1) & & \vdots \\ 0 & R(1) & R(0) & \ddots & 0 \\ \vdots & & \ddots & \ddots & R(-1) \\ 0 & \cdots & 0 & R(1) & R(0) \end{bmatrix}, \tag{13}$$

$$W = diag\left([A_1, A_2, \ldots, A_K, \ldots, A_1, A_2, \ldots, A_K]\right) \tag{14}$$

B is a vector which contains all the data of the K users. It is defined by:

$$B = \left[b_1^{(1)}, \ldots, b_K^{(1)}, b_1^{(2)}, \ldots, b_K^{(2)}, \ldots, b_1^{(N)}, \ldots, b_K^{(N)}\right]^T \tag{15}$$

And $\aleph$ is the vector of noise of length NK.

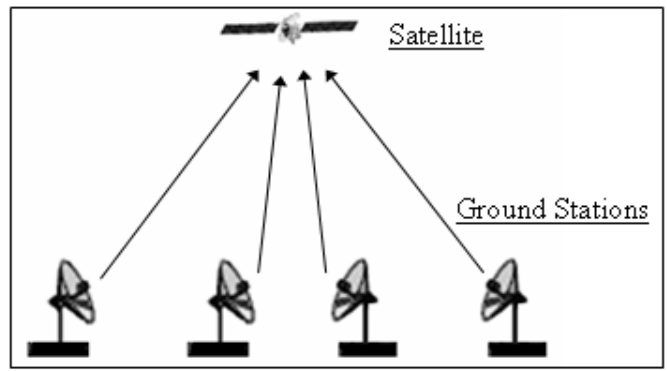

Figure 1. Communication via satellite uplink

## 4. CHANNEL ESTIMATION

In the uplink of a wireless system, the accesses of users are asynchronous and independent. Thus, each user signal is characterized by a specific propagation delay $\tau_k$. The channel estimation technique determines the amplitude and propagation delay of each user signal from the received signal presented in (1).

We consider that the propagation delays are multiple of the chip duration $T_c$, and $\forall k \neq l, \tau_k \neq \tau_l$. Thereafter, we consider that the chip period is a measurement unit for the propagation delays, and then, $\tau_k \in \{0,1,\ldots,N_c-1\}$. The estimation approach adopted here is to exploit the received signal part which suffers the least from multiple access interference. From expression (1) and (2), and figure (2), it is clear that the first bit of each user (except the $K^{th}$) is the least affected by the MAI compared to the other bits of the same user. So, due to propagation delays of users, we can exploit the following signal part:

$$\breve{r}(t) = \{r(t)/t \in [0, 2T_b]\} \quad (16)$$

Thus, the two first bits of each user signal can be considered as pilot symbols and are taken to $\{+1,+1\}$. For each user k, the bits $b_k^{(0)} = b_k^{(1)} = 1$, and, they can be omitted in the following discrete presentations.

Thereafter, the vector $\breve{r}$ which designates the received signal $\breve{r}(t)$ vector after sampling at a cadence of $1/T_c$, during two bit periods, can be written as:

$$\breve{r} = \sum_{k=1}^{K} \breve{r}_k + n \quad (17)$$

Where: $\breve{r}_k$ is the contribution of user k, it is expressed by:

$$\breve{r}_k = A_k \times \left[ O_{\tau_k}^T, c_k^{(0)}, \ldots, c_k^{(N_c-1)}, c_k^{(0)}, \ldots, c_k^{(N_c-1-\tau_k)} \right]^T \quad (18)$$

With: $O_{\tau_k}$ is the zeros column vector of length $\tau_k$.

$n$ is a $(2N_c \times 1)$ vector of noise resulting of $n(t)$ sampling during $2T_b$.

Estimation of signals amplitudes and propagation delays are computed alternatively according to the following algorithm:

*For k=1..K*

    *Determinate $\hat{\tau}_k$;*

    *Determinate $\hat{A}_k$;*

*end*

As the spreading codes are supposed known by the receiver, the propagation delays estimation is done by maximizing the cross-correlation between the users' signatures and the received signal part, after interference cancellation, as follows:

*For k=2..K*

$$\hat{r} = \breve{r} - I_{k-1} \quad (19)$$

$$\hat{\tau}_k = \arg \max_{\tau \in [\tau_{min}, \tau_{max}]} \left| s_k^T \hat{r}[\tau] \right|; k = 2..K \quad (20)$$

*end*

Where: $I_{k-1} = \hat{A}_{k-1} \times \left[ O_{\hat{\tau}_{k-1}}^T, c_{k-1}^{(0)}, \ldots, c_{k-1}^{(N_c-1)}, c_{k-1}^{(0)}, \ldots, c_{k-1}^{(N_c-1-\hat{\tau}_{k-1})} \right]^T \quad (21)$

and $\hat{r}[\tau] = \left[ \hat{r}^{(1+\tau)}, \hat{r}^{(2+\tau)}, \ldots, \hat{r}^{(N_c+\tau)} \right]^T \quad (22)$

With $\hat{r}^{(i)}$ is the i$^{th}$ element of $\hat{r}$. $\tau_{min}$ and $\tau_{max}$, which are expressed in $T_c$, are equal to 0 and $N_c - 1$ respectively. Estimation of the signals amplitudes is done successively from the least delayed signal to the most delayed. This estimation architecture is done with successive interference cancellation. The estimation of the first user signal amplitude is given by:

$$\hat{A}_1 = \frac{N_c}{N_s} \sum_{j=1}^{N_s} r_s^{(j)} v_1^{(j)} \qquad (23)$$

Where the column-vectors $r_s$ and $v_1$ are respectively $\breve{r}(t)$ and $s_1(t)$, after sampling at $N_s / T_c$ cadence, and $N_s$ is the number of samples per chip period. Thus, for a user k, $v_k$, whose length is $N_s N_c$, can be expressed as:

$$v_k = \left[ c_k^{(0)}, \ldots, c_k^{(0)}, c_k^{(1)}, \ldots, c_k^{(1)}, \ldots, c_k^{(N_c-1)}, \ldots, c_k^{(N_c-1)} \right]^T \qquad (24)$$

The estimation of signals amplitudes of the K-2 following users, $(k = 2..K-1)$, is computed by the following expression:

$$\hat{A}_k = \frac{N_c}{N_s} \sum_{j=\hat{\tau}_k N_s+1}^{(\hat{\tau}_k+1)N_s} \left( r_s^{(j)} - \sum_{l=1}^{k-1} \hat{e}_l^{(j)} \right) v_k^{(j)} \qquad (25)$$

For the user K, (the most delayed user), we will have:

$$\hat{A}_K = \frac{N_c}{(N_c - \hat{\tau}_K) N_s} \sum_{j=\hat{\tau}_K N_s+1}^{N_c N_s} \left( r_s^{(j)} - \sum_{l=1}^{K-1} \hat{e}_l^{(j)} \right) v_K^{(j)} \qquad (26)$$

Where: $\hat{e}_l^{(j)}$ is the j$^{th}$ element of $\hat{e}_l$, which is given by:

$$\hat{e}_l = \hat{A}_l \times \left[ v_l^T, v_l^T \right]^T \qquad (27)$$

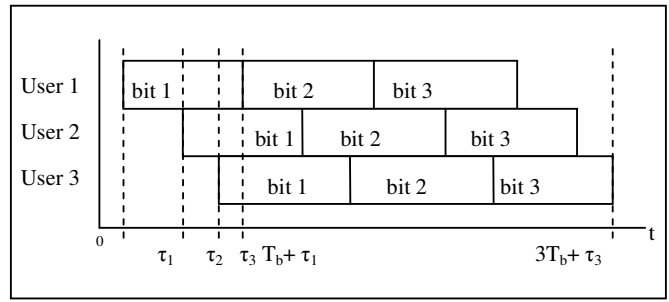

Figure 2. Transmission over asynchronous channel

## 5. MULTIUSER DETECTION

In this section, we develop a single stage SIC multiuser detection with integration of the channel estimation method presented above. Amplitudes and propagation delays of the signals, which are fed by the channel estimator, help to construct and subtract the interference from the received signal in order to recover the original data of each user by detecting them from an interference-cleaned signal. The detector is based on a single stage whose architecture is presented in figure 4, in which, the detections of bits are done successively. This detector is consisted of K interference cancellation unit (ICU) for the K users, (see figure 3).

The i$^{th}$ bits of users are serially detected and subtracted from the least delayed signal to the most delayed. In order to detect the i$^{th}$ bits of the K users, the receiver operates the part of the signal $r(t)$; $\tilde{r}^{(i)}(t)$, whose timing support is $[iT_b, (i+1)T_b + \hat{\tau}_K]$, with $i \in \{0, \ldots, N-1\}$. Thus, $\tilde{r}^{(i)}(t)$ can be expressed as:

$$\tilde{r}^{(i)}(t) = \prod_{[iT_b,(i+1)T_b+\hat{\tau}_K]}(t).r(t) \tag{28}$$

Where: $\prod_{[t_1,t_2]}(t)$ is the rectangular function defined by:

$$\prod_{[t_1,t_2]}(t) = \begin{cases} 1, & \text{if } t_1 \leq t \leq t_2 \\ 0, & \text{if not} \end{cases} \tag{29}$$

This function is used here as a temporal filtering operation.

At the input of the first ICU, the signal $z_0^{(i)}(t)$ is defined as follows:

$$z_0^{(i)}(t) = \tilde{r}^{(i)}(t) - \prod_{[iT_b,(i+1)T_b+\hat{\tau}_K]}(t).\Delta^{(i-1)}(t) \tag{30}$$

With, $\Delta^{(i-1)}(t)$ is the interference produced by the (i-1)$^{th}$ bits

of the K-1 users. It can be written as:

$$\Delta^{(i-1)}(t) = \sum_{k=2}^{K} \Delta_k^{(i-1)}(t) \tag{31}$$

Where, we have for a bit i:

$$\Delta_k^{(i)}(t) = \hat{A}_k \hat{b}_k^{(i)} s_k(t - iT_b - \hat{\tau}_k) \tag{32}$$

Now, we focus on the operation of the k$^{th}$ ICU. The signal used to detect the i$^{th}$ bit of user k is:

$$\tilde{r}_k^{(i)}(t) = \prod_{[iT_b+\hat{\tau}_k,(i+1)T_b+\hat{\tau}_k]}(t).z_{k-1}^{(i)}(t) \tag{33}$$

Thus, the detection is computed as follows:

$$\hat{b}_k^{(i)} = \text{sgn}\left[\int_{iT_b+\hat{\tau}_k}^{(i+1)T_b+\hat{\tau}_k} \tilde{r}_k^{(i)}(t).s_k(t - iT_b - \hat{\tau}_k)dt\right] \tag{34}$$

Where *sgn* denotes the sign function.

The output of the k$^{th}$ ICU is:

$$z_k^{(i)}(t) = z_{k-1}^{(i)}(t) - \prod_{[iT_b,(i+1)T_b+\hat{\tau}_K]}(t).\Delta^{(i)}(t) \tag{35}$$

This output feeds the input of the (k+1)$^{th}$ ICU and so on.

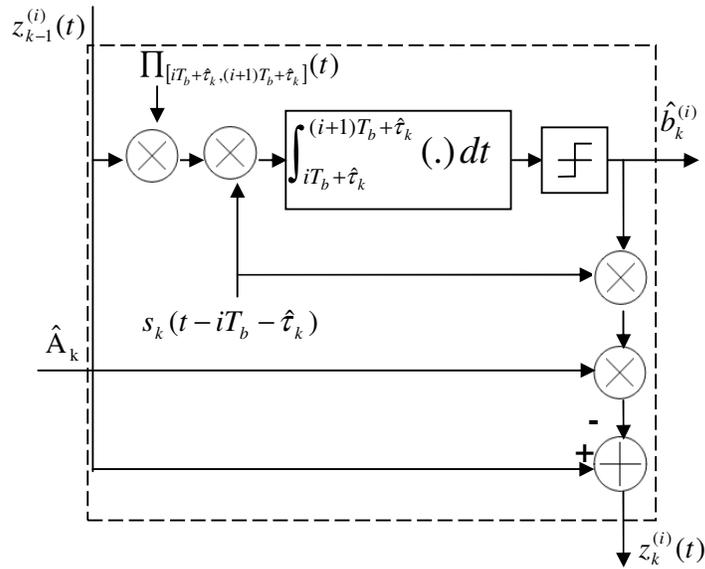

Figure 3. The k$^{th}$ Interference Cancellation Unit

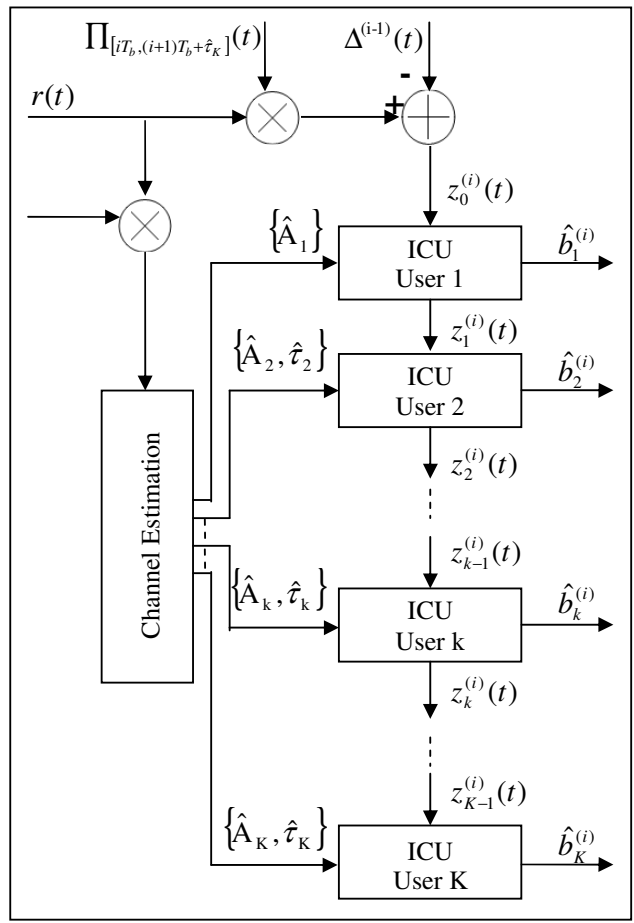

Figure 4. Receiver architecture

## 6. SIMULATION RESULTS

In order to evaluate the performance of the channel estimator and the SIC detector, we have simulated an asynchronous system with K=4 active users. Gold sequences of length 31 extended of one bit equal to 0 are used for spreading. The carrier of baseband signal is BPSK modulated. The wireless channel adopted here is an AWGN one, where each user signal has random delay propagation less than a symbol period and multiple of a chip period.

In figures 5 and 6, we have evaluated the channel estimation technique under near-far effect. When, signals amplitudes are different, they are arranged from the straightest to the weakest. In the first figure, we considered that the propagation delays are known by the receiver, and we have presented the power estimation error which is defined by:

$$PEE = \frac{1}{K} \sum_{k=1}^{K} \left| A_k - \hat{A}_k \right|^2 \qquad (36)$$

Although slightly degradation when near-far effect is presented, amplitude estimation shows good performance. For example for SNR=10 dB, the error for amplitude estimation is less than 0.1 when there is not a near-far problem. The bias of the estimator introduced with the near-far consideration is due to noise amplification. In fact, when we will add the white noise to the received signal, the power of the noise depends on the strongest signal, because the amplitude of the last presents practically the amplitude of the received signal. The propagation delay estimation of the system is improved with different signals amplitudes for high SNRs. This is due to interference suppression, since the more powerful signals are suppressed from the received signal. In fact, when we suppress powerful signals from the received signal, the maximization of the cross correlation becomes more precise, and then the error of propagation delays estimation decreases. For low SNRs, the results without near-far problem outperform those where near-far effect is presented. As a conclusion, near-far effect doesn't introduce great degradation in the system. Moreover, in the satellite communications, near-far effect is generally neglected.

In figure 7, amplitude estimation is evaluated with different values of Ns. The results depicted show that increasing of the number of samples per chip improves significantly amplitude estimation quality. This is due to the refinement of the estimation. More the number of samples increase more the estimation quality is improved. Secondly, increasing of sampling cadence leads to time computation increasing.

In figure 8, we have evaluated the channel estimation method (amplitudes and propagation delays conjointly) with Ns=120 and equal signals amplitudes. Average amplitude estimation error ($1/K \sum_{k=1}^{K} \left| A_k - \hat{A}_k \right|$) is presented. The low values of estimation error obtained and the slight differences between the two curves show the robustness of the channel estimation technique and particularly the good performance of the propagation delay estimation.

The channel estimation technique is evaluated also with the single stage SIC detector evaluation. In figure 9, we have evaluated and compared the SIC detector with two implementation approaches. The first implementation is that developed in this paper. And, in the second approach, we have combined the SIC with the Matched filter detector (i.e. the conventional one) in order to suppress the interference of the $(i+1)^{th}$ bits while detecting the $i^{th}$ bits. This detector is mentioned by SIC/MF in the figure 9. So, we have considered for the SIC/MF that (see figure 4):

$$z_0^{(i)}(t) = \tilde{r}^{(i)}(t) - \prod_{[iT_b,(i+1)T_b+\hat{\tau}_K]}(t).\left(\Delta^{(i-1)}(t) + \Delta^{(i+1)}(t)\right) \quad (37)$$

The different average Bit Error Ratios (BER) presented in figure 9 show that the single stage SIC receiver has good performance and outperforms significantly the single stage SIC/MF detector. The explanation of that result is due to the poor detection quality of the matched filter method which is integrated into the SIC/MF detection technique. In fact, bad detection of data will amplify the interferences and not subtract them when the operation of interference cancellation will be done. The results depicted in figure 9 show also the good quality of the channel estimator since the performances of the detector with channel estimation and those with channel knowledge are very similar.

## 7. CONCLUSIONS

In this paper, a channel estimation technique and a single stage multiuser detector for asynchronous wireless communications are presented. They are based on signals correlation and interference suppression. The channel estimation method takes into account the timing offsets between the signals, as the system is asynchronous, and exploits consequently the beginning of the total received signal, which suffers the least from multiuser interference. The signals amplitudes and propagation delays are estimated successively for each active user in the uplink of the satellite system. The estimated channels parameters are then fed to the SIC detector to recover the original data of users. The performances of the channel estimation technique and the multiuser detector which implements it are evaluated through computer simulations. In some ones, we have considered a near-far environment. Due to the asynchronous character of the channel, the detector performances are closely dependent on channel estimator ones, and especially the propagation delays estimation performance. The obtained results show good performance and computation simplicity of the channel estimation method and the multiuser detection technique. As a future work, it will be interesting to deal with channel estimation and multiuser detection with use of other multiple access methods and integration of FEC coding, and consider mulitbeam technology for wireless systems.

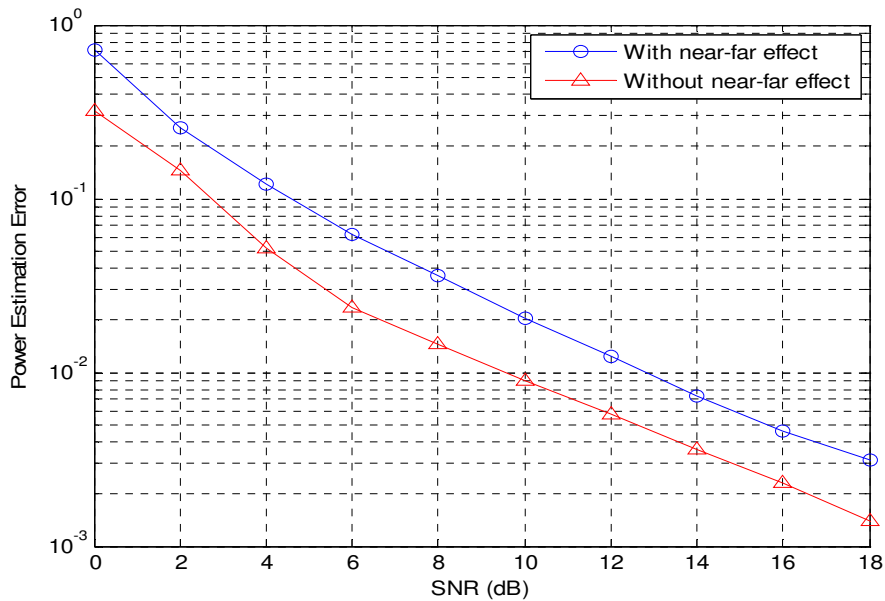

Figure 5. Performance of power estimation under near-far effect

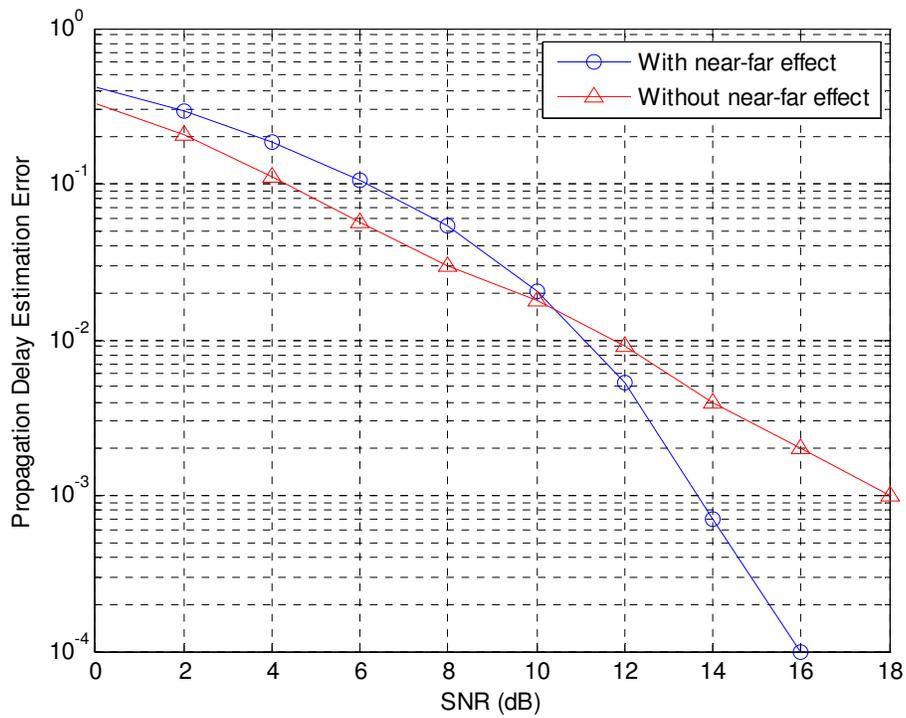

Figure 6. Performance of propagation delay estimation under near-far effect

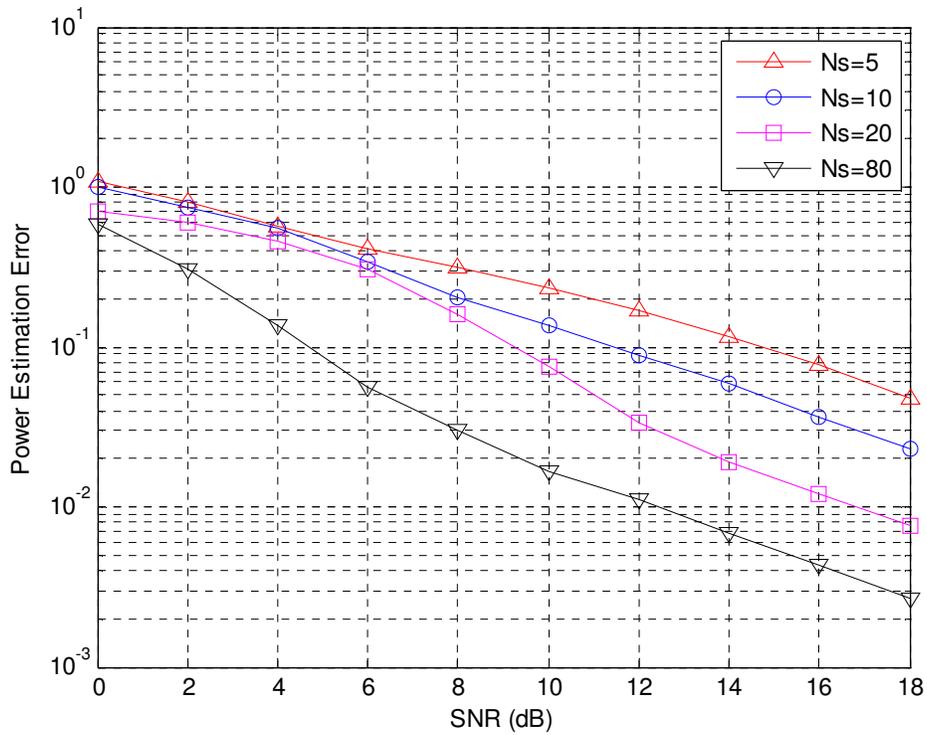

Figure 7. Performance of power estimation with different values of Ns

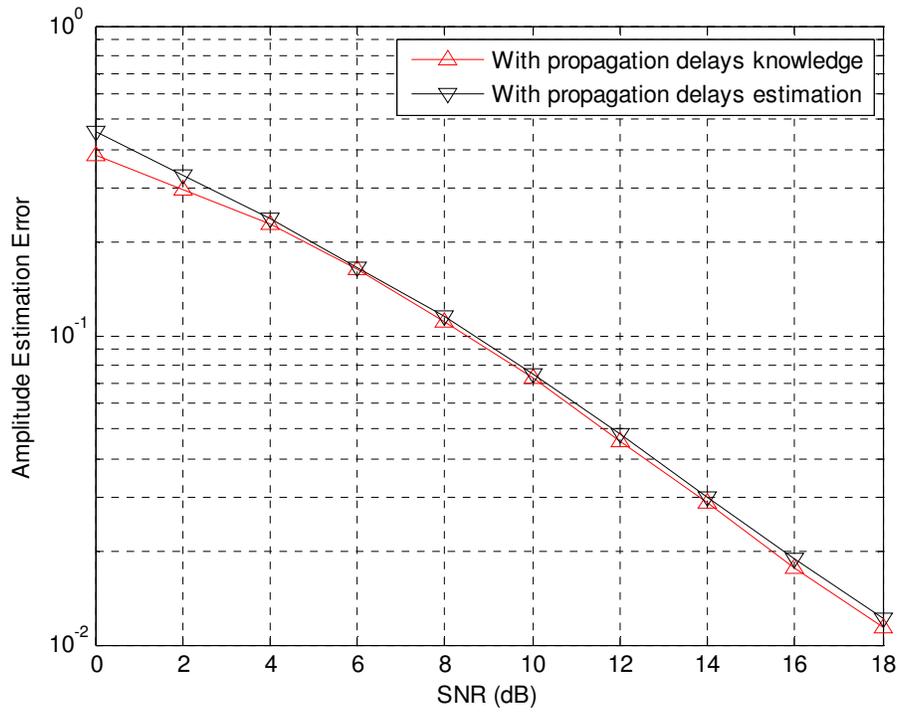

Figure 8. Performance of amplitude estimation with propagation delay estimation

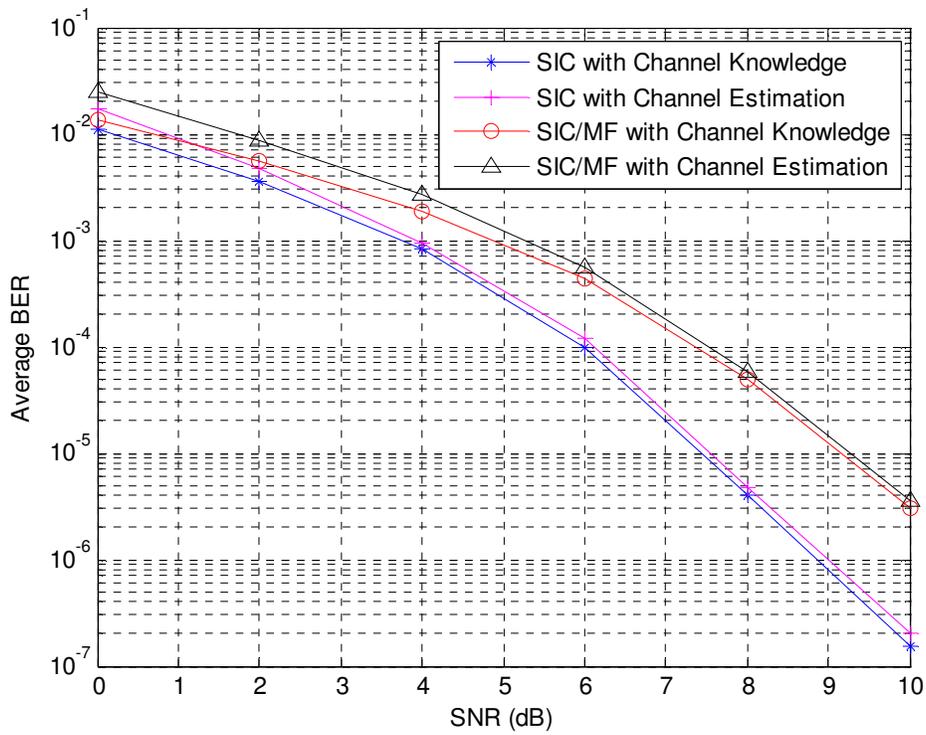

Figure 9. Performance of SIC detector with channel estimation

## REFERENCES

[1]   S. Verdu, Multiuser Detection, Cambridge University Press, 1998.

[2]   V. Kekatos, A. A.Rontogiannis, and K. Berberidis, "A robust parametric technique for multipath channel estimation in the uplink of a DS-CDMA system", EURASIP Journal on Wireless Communications and Networking, volume 2006, Article ID 47938, pages 1-12.

[3]   S. Verdu, "Minimum probability of error for asynchronous Gaussian multiple access channels", IEEE Trans. Inf. Theory, vol. IT-32, no. 1, pp.85-96, Jan. 1986.

[4]   H. Hachaichi, H. Chaouech and R. Bouallegue, "Multi-path channel estimation methods for UMTS TDD", 11th Communications and Networking Simulation Symposium (CNS 2008), Ottawa, Canada, April 14-17 2008.

[5]   M. Varanasi and B. Aazhang, "Multi-stage detection in asynchronous code-division multiple access communications", IEEE Trans. Commu., vol. 38, pp. 509-519, Apr. 1990.

[6]   J. P. Millerioux, "Multiuser detection techniques for multibeam satellite communications", Ph.D dissertation, Telecom ParisTech, France, 2006.

[7]   M. Sirbu, "Channel and delay estimation algorithms for wireless communication systems", thesis of Helsinki University of Technology (Espoo, Finland), December 2003.

[8]   A. Amer and F. Gebali, "General model for infrastructure multichannel wireless LANs", International Journal of Computer Networks & Communications (IJCNC), Vol.2, No.3, May 2010.

[9]   M. El-Tarhuni, M. Hassan and A. Ben Sediq, "A joint power allocation and adaptive channel coding scheme for image transmission over wireless channels", International Journal of Computer Networks & Communications (IJCNC), Vol.2, No.3, May 2010.

[10]  M. Hassan, T. Landolsi and H. Mukhtar, "A channel-aware and occupancy-dependant scheduler for video transmission over wireless channels", International Journal of Computer Networks & Communications (IJCNC), Vol.2, No.5, September 2010.

[11]  H. Chaouech and R. Bouallegue, "Channel estimation and detection for multibeam satellite communications", accepted to be published in IEEE Asia Pacific Conference on circuits and Systems, Kuala Lumpur, Malaysia, December 6-9 2010.

[12]  K. Khairnar and S. Nema, "Comparison of multiuser detectors of DS CDMA system", World Academy of Science, engineering and Technology 2005.

[13]  R. Lupas and S. Verdu, "Near-far resistance of multiuser detectors in asynchronous channels", IEEE Transactions on Communications, Vol.38, No.4, April 1990.

[14]  S. S. H. Wijayasuriya, G. H. Norton and J. P. McGeehan, "A sliding window decorrelating receiver for multiuser DS-CDMA mobile radio networks", IEEE Transactions on Vehicular Technology, Vol. 45, No. 3, August 1996.

[15]  B. F. Beidas, H. El Gamal and S. Kay, "Iterative interference cancellation for high spectral efficiency satellite communications", IEEE Transactions on Communications, Vol. 50, No. 1, January 2002.

[16]  L. C. Zhong, Z. Siveski, R. E. Kamel and N. Ansari, "Adaptive multiuser CDMA detector for asynchronous AWGN channels-steady state and transient analysis", IEEE Transactions on Communications, Vol. 48, No. 9, September 2000.

[17]  K. Ko, M. Joo, H. Lee and D. Hong, "Performance analysis for multistage interference cancellers in asynchronous DS-CDMA systems", IEEE Communications Letters, Vol. 6, No. 12, December 2002.

[18]  S. H. Han and J. H. Lee, "Multi-stage partial parallel interference cancellation receivers for multi rate DS-CDMA systems", IEEE Transactions on Communications, Vol. E86-B, No. 1, January 2003.

[19]     J. P. Millerioux, M. L. Boucheret, C. Bazile and A. Dicasse, "Frequency estimation in iterative interference cancellation applied to multibeam satellite systems", EURASIP Journal on Wireless Communications and Networking, Volume 2007, Article ID 62310, 12 pages.

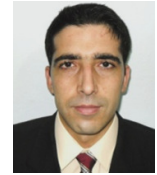

**Helmi CHAOUECH** received the engineering degree in telecommunications in 2006 and the M.S degree in communications systems in 2007 from the National Engineering School of Tunis, Tunisia. Currently, he is with the 6'Tel Research Unit, Higher School of Communications of Tunis (SUP'COM), Tunisia, as a Ph.D student and hi is an assistant in the Faculty of Economic Sciences and Management of Nabeul, Tunisia. His research interests include channel estimation, multiuser detection, wireless communication theory and multibeam satellite communications.

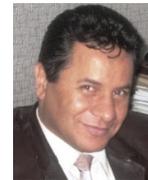

**Ridha BOUALLEGUE** received the Ph.D degree in electronic engineering from the National Engineering School of Tunis in 1998. In Mars 2003, he received the HDR degree in multiuser detection in wireless communications. From 2005 to 2008, he was the Director of the National Engineering School of Sousse, Tunisia. In 2006, he was a member of the national committee of science technology. Since 2005, he was the Director of the 6'Tel Research Unit in SUP'COM. Currently, hi is the Director of Higher School of Technology and Informatique, Tunis, Tunisia. His current research interests include wireless and mobile communications, OFDM, space-time processing for wireless systems, multiuser detection, wireless multimedia communications, and CDMA systems.